\documentclass[nofootinbib,reprint,aps,prb,twocolumn,floatfix]{revtex4-2}
\usepackage{amsmath}
\usepackage{amsfonts}
\usepackage{braket}
\usepackage[dvipsnames]{xcolor}
\usepackage{graphicx}
\DeclareGraphicsExtensions{.pdf, .png}
\graphicspath{ {./images/} }
\usepackage{hyperref}
\usepackage{soul}

\newcommand{\kay}[0]{\mathbf{k}}
\newcommand{\ar}[0]{\mathbf{r}}
\newcommand{\ee}[0]{\mathbf{e}}
\newcommand{\Bee}[0]{\mathbf{B}}
\newcommand{\en}[0]{\mathbf{n}}
\newcommand{\Em}[0]{\tilde{M}}
\newcommand{\Oh}[0]{\mathcal{O}}
\newcommand{\Ay}[0]{\mathbf{A}}

\begin{document}

\title{Surface theory of a second-order topological insulator beyond the Dirac approximation}

\author{B.~A.\ Levitan}
\email{levitanb@physics.mcgill.ca}
\author{L. Goutte}
\author{T.\ Pereg-Barnea}
\affiliation{Department of Physics and the Centre for the Physics of Materials, McGill University, 3600 rue University, Montr\'{e}al, Quebec H3A 2T8, Canada}

\begin{abstract}
We study the surface states and chiral hinge states of a 3D second-order topological insulator in the presence of an external magnetic gauge field. Surfaces pierced by flux host Landau levels, while surfaces parallel to the applied field are not significantly affected. The chiral hinge modes mediate spectral flow between neighbouring surfaces.  As the magnetic field strength is increased, the surface Landau quantization deviates from that of a massive Dirac cone. Quantitatively, the $n = 0$ Landau level falls inside the surface Dirac gap, and not at the gap edge. The $n \ne 0$ levels exhibit a further, qualitative discrepancy: while the massive Dirac cone is expected to produce pairs of levels ($\pm n$) which are symmetric around zero energy, the $n$ and $-n$ levels become asymmetric in our lattice model --- one of the pair may even be absent from the spectrum, or hybridized with the continuum. In order to resolve the issue, we extend the standard 2D massive Dirac surface theory, by including additional Hamiltonian terms at $\Oh(\kay^2)$. While these terms do not break particle-hole symmetry in the absence of magnetic field, they lead to the aforementioned Landau level asymmetry once the magnetic field is applied.  
We argue that similar $\Oh(\kay^2)$ correction terms are generically expected in lattice models containing gapped Dirac fermions, using the BHZ model of a 2D topological insulator as an example.
\end{abstract}

\maketitle

\section{Introduction}
The majority of symmetry-protected topological phases (SPTs) are completely characterized by their bulk-boundary correspondence \cite{Schnyder:2008classification, Kitaev:2009periodictable, Ryu:2010tenfoldway, KaneMele:Z2SpinHall, FuKaneMele:3DTI, FuKane:InversionTI,Hatsugai:ChernEdge,Hasan_Moore:2011Review, Hasan:Colloquium}. A typical bulk-boundary correspondence states that a bulk topological invariant --- often computed from a band structure, and taking values in $\mathbb{Z}$ or $\mathbb{Z}_n$ --- counts the number of protected lower-dimensional gapless modes, which arise at appropriate surface terminations. In the case of a topological insulator \cite{KaneMele:Z2SpinHall, BernevigHughesZhang, FuKaneMele:3DTI, FuKane:InversionTI,Hatsugai:ChernEdge,Hasan_Moore:2011Review, Hasan:Colloquium}, the bulk is by definition fully gapped, and so the protected gapless modes dominate the low-energy physics.

Accordingly, when studying a given SPT, it is often productive to write down a theory describing only the boundary modes. In the case of a 3D strong (first-order) topological insulator, the boundary theory describes an unusual 2D metal, with energy bands crossing linearly at an odd number of Dirac points. In the case of a 3D second-order topological insulator (SOTI) \cite{Benalcazar:MultipoleInsulators, Benalcazar:Pumping, Langbehn:2017Reflection, Song:2017Rotation, Geier:2018OrderTwo, Khalaf:2018Inversion, Schindler:HOTI, Khalaf:2018Symmetry, Schindler:2018Bismuth, Slager:2015Impurities}, the protected metal is confined to $3-2=1$D regions of the boundary. For the $C_4^z \mathcal{T}$-symmetric chiral SOTI \cite{Schindler:HOTI} (where $C_4^z$ is a $\pi/2$ rotation around the $z$ axis and $\mathcal{T}$ is time-reversal), the 1D modes reside along ``hinges" where surfaces $\perp \ee_x$ meet surfaces $\perp \ee_y$.

Like its first-order cousin, the surface of a chiral SOTI can be productively described in the language of Dirac fermions. The key difference is that the surface Dirac fermion acquires a mass gap, which alternates in sign between one surface component and its neighbour. This sign change can be thought of as a topological defect, which pins a massless 1D Jackiw-Rebbi mode \cite{JackiwRebbi} running along the hinge --- we generalize the Jackiw-Rebbi calculation to include a magnetic gauge field in subsection \ref{subsec:hinge_theory}. The key topological features of the SOTI surface are thus captured by a 2D massive Dirac model. 

Unlike the first-order case where the 2D surface state is gapless, the SOTI surface gap implies that dispersion near the ``massive Dirac point" is quadratic in momentum rather than linear. This is the first hint that often-overlooked Hamiltonian terms at $\Oh(\kay^2)$ should be included in a full leading-order analysis. Close to the Dirac point, such terms have only a small effect on the spectrum, justifying the standard (Dirac) picture. Surprisingly, the $\Oh(\kay^2)$ terms take on qualitative significance in the presence of a strong magnetic field. In subsection \ref{subsec:landau_k2_corrections}, we show how the $\Oh(\kay^2)$ terms explain the deviations from relativistic Landau quantization we reported in a previous numerical study \cite{Levitan:SOTImagnetotransport}. In addition to a nonlinear modification of the level spacing, the entire Landau spectrum experiences a shift proportional to the applied magnetic field $B$. Consequently, the lowest Landau level appears \textit{inside} the surface Dirac gap, and the $n \ne 0$ levels are asymmetric around zero energy.

Using the BHZ model of a 2D topological insulator as an example, we argue that similar $\Oh (\kay^2)$ terms will appear in \textit{any} lattice model containing a massive Dirac fermion. Therefore, our results should qualitatively generalize to a variety of other settings. 

\section{Lattice model}
We begin by considering a minimal four-band tight-binding model on the simple cubic lattice, with a topologically-nontrivial chiral hinge insulator phase protected by $C_4^z \mathcal{T}$-symmetry (the product of a fourfold-rotation and time-reversal) \cite{Schindler:HOTI, Levitan:SOTImagnetotransport}. We introduce a magnetic field via Peierls substitution \cite{Hofstadter:butterfly, Levitan:SOTImagnetotransport}: the gauge field $\Ay$ associates a $U(1)$ Peierls phase factor to each hopping process on the lattice, breaking $C_4^z \mathcal{T}$. In units where the lattice constant $a$ is set to $1$, the Hamiltonian is
\begin{widetext}
\begin{equation}	\label{eq:hamiltonian}
	H = \sum_{\ar} \Bigg \lbrace c^{\dagger}_{\ar} \left[ M \hat{\sigma}_0 \hat{\tau}_z  \right] c_{\ar}	\\
		+ \sum_{j = x,y,z} \left( c^{\dagger}_{\ar + \ee_j} 
			\left[ \frac{e^{i \theta_{\ar}^{j}}}{2} \left(
			- t \hat{\sigma}_0 \hat{\tau}_z + i \Delta_1 \hat{\sigma}_j \hat{\tau}_x
			+ \Delta_{2, j} \hat{\sigma}_0 \hat{\tau}_y 
			\right) \right] c_{\ar} + \mathrm{h.c.} \right)
		\Bigg \rbrace,
\end{equation}
\end{widetext}
where $\Delta_{2,x} = - \Delta_{2,y} = \Delta_2$, and $\Delta_{2,z} = 0$. $c_{\ar}$ is a four-component fermion annihilation operator at lattice site $\ar$. $\hat{\sigma}_j$ and $\hat{\tau}_j$ are the $2 \times 2$ Pauli matrices, acting on the spin ($\uparrow, \downarrow$) and orbital ($0, 1$) degrees of freedom respectively --- throughout, we use hats ({ \large $_{\hat{}}$}) to denote matrices acting in the spin $\otimes$ orbital space. The Peierls phases are given by $e^{i \theta_{\ar}^{j}} = e^{i (q / \hbar) \int_{\ar}^{\ar + \ee_j} \mathrm{d} \boldsymbol{\ell} \cdot \Ay}$, with the electron charge $q = -e$.

In addition to the Peierls phases, which affect the motion of fermions on the lattice, a magnetic field would also generally involve a Zeeman term $\propto \boldsymbol{B} \cdot \boldsymbol{\sigma}$. We showed in a previous work \cite{Levitan:SOTImagnetotransport} that the Zeeman effect shifts the surface Dirac masses, but has no other significant impact. We therefore neglect the Zeeman effect in our present analysis, focusing on the richness of the orbital effects resulting from an applied magnetic field.

Without the magnetic field (i.e.~when $\theta_{\ar}^{j} = 0$), for $\Delta_1 \Delta_2 \ne 0$, the bulk gap of the Hamiltonian of Eq.~\eqref{eq:hamiltonian} can only close at the $C_4^z \mathcal{T}$-invariant momenta $\boldsymbol{\Gamma} = (0, 0, 0)$, $\boldsymbol{X} = (0, 0, \pi)$, $\boldsymbol{M} = (\pi, \pi, 0)$ and $\boldsymbol{R} = (\pi, \pi, \pi)$. The energy eigenvalues at those momenta are
\begin{subequations}
	\begin{equation}
		E_{\text{bulk}}^{\pm} (\boldsymbol{\Gamma}) = \pm (M - 3 t)
	\end{equation}
	\begin{equation}
		E_{\text{bulk}}^{\pm} (\boldsymbol{X}) = \pm (M - t)
	\end{equation}
	\begin{equation}
		E_{\text{bulk}}^{\pm} (\boldsymbol{M}) = \pm (M + t)
	\end{equation}
	\begin{equation}
		E_{\text{bulk}}^{\pm} (\boldsymbol{R}) = \pm (M + 3 t)
	\end{equation}
\end{subequations}
where each energy is doubly degenerate.
When $1 < | M / t | < 3$ and $\Delta_1 \ne 0 = \Delta_2$, the model is in a (strong, first-order) topological insulator phase, protected by $\mathcal{T}$. Turning on the $\Delta_2$ term moves the model from the first- to the second-order topological phase. This term breaks $\mathcal{T}$ while preserving $C_4^z \mathcal{T}$, gapping out the Dirac electrons on surfaces perpendicular to $\ee_{x}$ or $\ee_y$. $C_4^z \mathcal{T}$ symmetry enforces that the sign of the surface Dirac mass alternates between one surface component (say, $\perp \ee_x$) and its neighbour (say, $\perp \ee_y$). Gapless chiral one-dimensional modes arise, pinned to the hinges along which the mass changes sign. We will assume throughout that parameters are chosen to fall within the second-order topological phase $1 < | M / t | < 3$, $\Delta_1 \Delta_2 \ne 0$, and we will take $M, t > 0$ in order to situate the surface massive Dirac cones at zero in-plane momentum ($\kay_{\parallel} = 0$, where, for example, $\kay_{\parallel} = (0, k_y, k_z)$ on the surface normal to $\ee_x$).

We consider a magnetic field along the $x$-direction, and use the Landau gauge $\Ay = (0, 0, B y)$. The Peierls phases are then $\theta_{\ar}^x = \theta_{\ar}^y = 0$, $\theta_{\ar}^{z} = - 2 \pi y \Phi / \Phi_0$, where $\Phi / \Phi_0$ is the magnetic flux-per-plaquette in the $yz$-plane measured in units of the flux quantum $\Phi_0 = 2 \pi \hbar / e$. This choice of gauge preserves the translation symmetries along $x$ and $z$ (but not along $y$), so the Hamiltonian of Eq.~\eqref{eq:hamiltonian} can be block-diagonalized over the momentum component $k_z$ (by Fourier transforming in the $z$ direction). Unless otherwise stated, our numerical data correspond to open boundary conditions in the $x$ and $y$ directions and periodic boundary conditions along $z$, with bulk parameters $M/t = 2.3$, $\Delta_1/t = 0.8$, and $\Delta_2/t = 0.5$ ($t > 0$).

\begin{figure*}
	\centering
	\includegraphics[width=\linewidth]{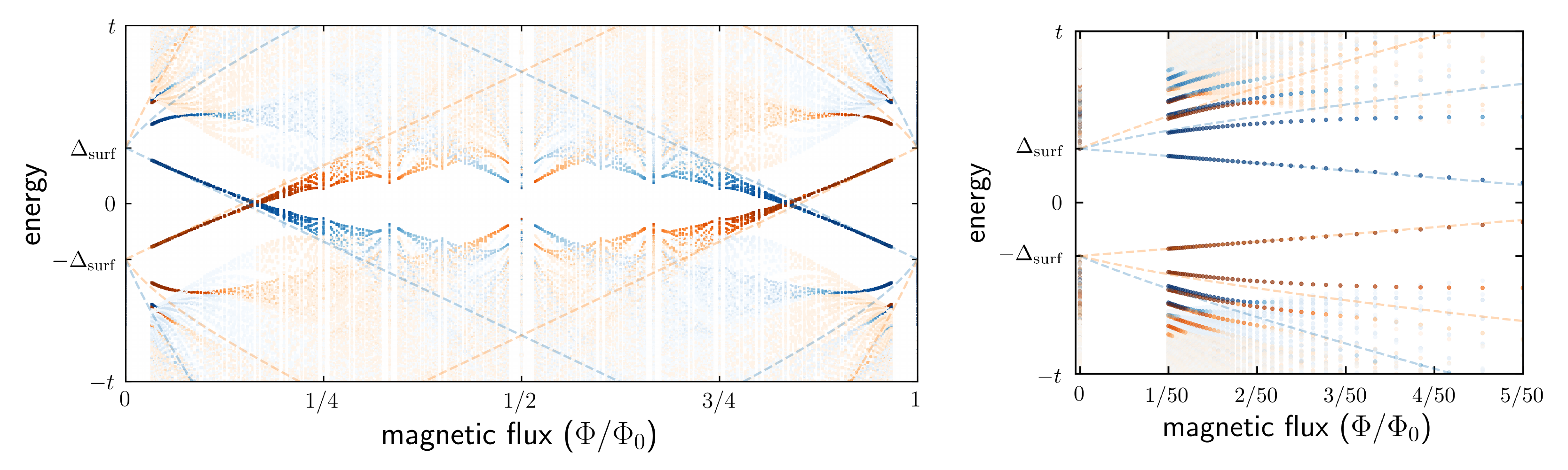}
	\caption{\label{fig:butterfly} At the surface of a 3D SOTI, the energy spectrum displays a complex dependence on applied magnetic flux-per-plaquette $\Phi$, evocative of Hofstadter's butterfly (left panel). The weak-field regime (right panel) is captured by a surface theory at $\Oh(\kay_{\parallel}^2)$,
		whose lowest Landau levels are shown as dashed lines. Numerical data correspond to an infinite slab geometry, with thickness $L = 30$ ($L = 50$) for the left (right) panel. Orange 	(blue) points correspond to states localized near the surface at $x = 0$ ($x = L-1$). Lightly-coloured points correspond to bulk states.}
\end{figure*}

When particles on a two-dimensional lattice are subject to a strong magnetic field, the physics of Hofstadter's butterfly \cite{Hofstadter:butterfly} is generically expected; the competition between magnetic and lattice length scales produces a fractal pattern in the energy spectrum as a function of flux per plaquette. The surface states of the chiral hinge insulator are effectively two-dimensional, and their spectrum as a function of flux, shown in Fig.~\ref{fig:butterfly}, accordingly resembles the classic Hofstadter result. It is important to note that the chiral hinge insulator phase does not generically require surface states to exist inside the bulk gap. The Hofstadter-like structure in Fig.~\ref{fig:butterfly} will be seen in regions of the topological phase where the surface gap is smaller than the bulk gap. Further, note that the Hofstadter butterfly is only observable when the magnetic flux per unit cell is an appreciable fraction of the flux quantum $\Phi_0$. For atomic lattices, with interatomic distances at the scale of $10^{-10}$ m, the magnetic field strength required to escape from the weak-field limit ($\Phi / \Phi_0 \ll 1$) is infeasibly large. Superlattice structures with significantly larger lattice constants may enable the observation of the Hofstadter butterfly, and will enhance all lattice-related effects, such as the quadratic corrections discussed in section \ref{sec:k2_corrections}.


%

\section{Boundary theory} \label{sec:boundary_theory}

\subsection{Surface Dirac fermion}	\label{subsec:surface_dirac}
At zero magnetic field and with periodic boundary conditions in all directions (sample volume $\mathcal{V} = L_x \times L_y \times L_z$), the microscopic tight-binding model of Eq.~\eqref{eq:hamiltonian} is conveniently represented in momentum space as $H = \sum_{\kay} c_{\kay}^{\dagger} \hat{\mathcal{H}} (\kay) c_{\kay}$, with $c_{\ar} = \frac{1}{\sqrt{\mathcal{V}}} \sum_{\kay} e^{i \kay \cdot \ar} c_{\kay}$. The Bloch Hamiltonian $\hat{\mathcal{H}} (\kay)$ has long-wavelength ($\kay \rightarrow 0$) limit
\begin{multline}	\label{eq:smallk_hamiltonian}
	\hat{\mathcal{H}} (\kay) = \left( \Em + \frac{t}{2} \kay^2 + \Oh(\kay^4) \right) \hat{\sigma}_0 \hat{\tau}_z	\\
		+ \Delta_1 \left(\kay \cdot \hat{\boldsymbol{\sigma}} + \Oh(\kay^3) \right) \hat{\tau}_x	\\
		- \frac{\Delta_2}{2} \left( k_x^2 - k_y^2 + \Oh (\kay^4) \right) \hat{\sigma}_0 \hat{\tau}_y
\end{multline}
where $\Em = M - 3t$. Recall that we work in units where $a = 1$, so that $\kay$ is effectively dimensionless; $M$, $t$, $\Delta_1$ and $\Delta_2$ have units of energy.

Consider a surface termination with outward-facing normal vector $\en = \pm \ee_j$ ($j \in \lbrace x, y \rbrace$). For the minimal model under consideration, surfaces normal to $\pm \ee_z$ would be gapless \cite{Schindler:HOTI}; we therefore assume periodicity along the $z$-direction for convenience. As we have described in a previous work \cite{Levitan:SOTImagnetotransport}, a massive Dirac fermion residing on the surface can be identified by assuming exponential decay in the $-\en = \mp \ee_j$ direction (into the bulk),
i.e.~by making the replacement $k_j \rightarrow \mp i \kappa$ in Eq.~\eqref{eq:smallk_hamiltonian}, with $\mathrm{Re} [\kappa]> 0$.

For each choice of surface $\en$, we can decompose the Hamiltonian into pieces describing motion parallel and perpendicular to the surface, keeping terms up to $\Oh(\kay^2)$. For example, at the $\en = +\ee_x$ surface:
\begin{subequations}
	\begin{equation}
		\hat{\mathcal{H}}^{(\ee_x)}_{\parallel} = \frac{t}{2} \kay_{\parallel}^2 \hat{\sigma}_0 \hat{\tau}_z + \Delta_1 (\kay_{\parallel} \cdot \hat{\boldsymbol{\sigma}}) \hat{\tau}_x 
			+ \frac{\Delta_{2}}{2} k_{y}^2 \hat{\sigma}_0 \hat{\tau}_y,
	\end{equation}
	\begin{multline}	\label{subeq:H_perp_x}
		\hat{\mathcal{H}}^{(\ee_x)}_{\perp} = \left( \Em - \frac{t}{2} \kappa^2 \right) \hat{\sigma}_0 \hat{\tau}_z - i \kappa \Delta_1 \hat{\sigma}_x \hat{\tau}_x	\\
			+ \frac{\Delta_{2}}{2} \kappa^2 \hat{\sigma}_0 \hat{\tau}_y.
	\end{multline}
\end{subequations}
$\kay_{\parallel}$ denotes the momentum in the plane of the surface; on the $\ee_x$ surface, $\kay_{\parallel} = (0, k_y, k_z)$. For a given surface $\en = \pm \ee_j$, at $\kay_{\parallel} = 0$, the effective Schr\"{o}dinger equation is $\hat{\mathcal{H}}^{(\pm \ee_j)}_{\perp} \ket{\Psi^{(\pm \ee_j)}} = E \ket{\Psi^{(\pm \ee_j)}}$.

Since $[\hat{\mathcal{H}}^{(\pm \ee_j)}_{\perp}, \hat{\sigma}_j] = 0$ ($j \in \lbrace x, y \rbrace$), we can organize the eigenstates of $\hat{\mathcal{H}}^{(\pm \ee_j)}_{\perp}$ according to their $\hat{\sigma}_j$ eigenvalue, $\ket{\Psi^{(\pm \ee_j)}_{\sigma_j}} \propto \ket{\sigma_j} \otimes \ket{\chi^{(\pm \ee_j)}_{\sigma_j}}$. $\lbrace \ket{\sigma_j = +1}, \, \ket{\sigma_j = -1}\rbrace$ are the eigenstates of $\hat{\sigma}_j$, labelled by their eigenvalue.
Following the calculation we described in Ref.~\cite{Levitan:SOTImagnetotransport}, we find the orbital components
\begin{equation}
	\ket{\chi^{(\ee_x)}_{\sigma_x}} \propto \begin{pmatrix}
		- i \chi_{\sigma_x}	\\
		1
	\end{pmatrix},
\end{equation}
with $\chi_{\sigma_x} = \frac{1}{t} \left( \Delta_2 + \sigma_x \sqrt{t^2 + \Delta_2^2} \right) \in \mathbb{R}$, meaning that $\ket{\chi^{(\ee_x)}_{\sigma_x}}$ lies in the $\tau_y \tau_z$-plane.
The energies are
\begin{equation}
	E^{(\ee_x)}_{\sigma_x} (k_y = k_z = 0) = - \Delta_{\text{surf}} \sigma_x \equiv \frac{\Em \Delta_2}{\sqrt{t^2 + \Delta_2^2}} \sigma_x.
\end{equation}
Recall that here, $\sigma_x = \pm 1$ is a c-number.

\begin{figure}
	\centering
	\includegraphics[width=0.9\linewidth]{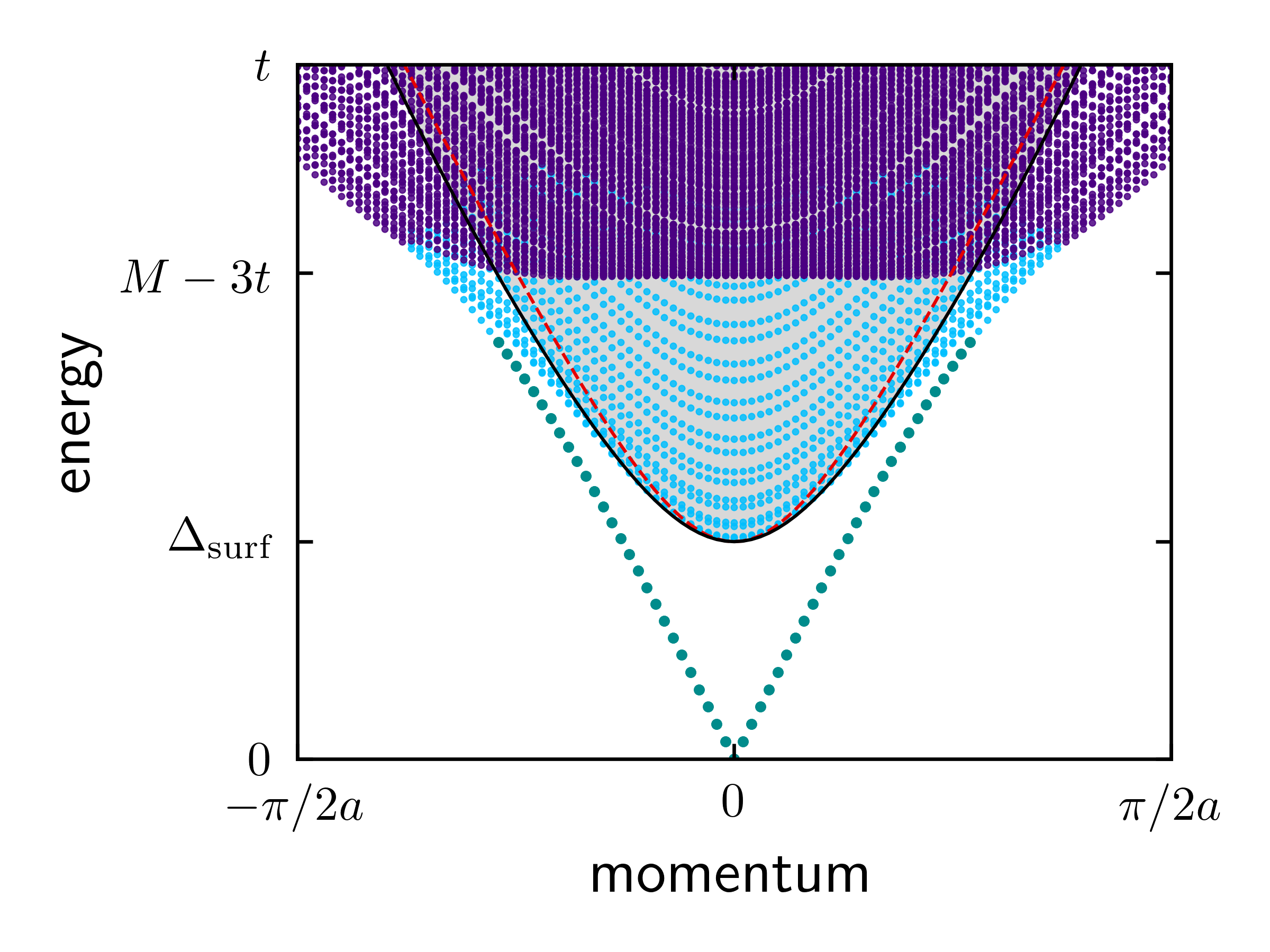}
	\caption{\label{fig:zerofield} Without an applied magnetic field, the spectrum of a 3D SOTI can be broken into bulk states (gapped, purple), surface states 
		(gapped, light blue), and hinge states (gapless, turquoise). Only positive-energy states are shown; the energy spectrum is symmetric around zero energy.
		A simple surface theory can be constructed which accurately predicts the surface mass gap $\Delta_{\text{surf}}$ and the associated dispersion at small in-plane momentum 
		$\kay_{\parallel}$ (see subsections \ref{subsec:surface_dirac} and \ref{subsec:zerofield_k2_corrections}). 
		The red dashed lines correspond to the massive Dirac Hamiltonian (at $k_y = 0$). 
		The black lines and grey shaded region include Hamiltonian terms at $\Oh(\kay_{\parallel}^2)$. 
		The $\Oh (\kay_{\parallel}^2)$ corrections become important in the presence of a strong magnetic field (see subsection \ref{subsec:landau_k2_corrections}).
		The sample is of cross-sectional dimensions $L_x \times L_y = 30 \times 30$, with periodic boundary conditions along the $z$ direction.}
		
\end{figure}

To obtain an approximate description of the long-wavelength surface states, we proceed in the spirit of $\kay \cdot \mathbf{p}$ perturbation theory and project $\hat{\mathcal{H}}_{\perp}^{(\ee_x)} + \hat{\mathcal{H}}_{\parallel} ^{(\ee_x)}$ onto the subspace spanned by $\lbrace \ket{ \Psi^{(\ee_x)}_{\sigma_x = +1}},  \ket{ \Psi^{(\ee_x)}_{\sigma_x = -1}} \rbrace$. For $\kay_{\parallel} \ll 1$,
\begin{multline}	\label{eq:surface_dirac}
	\hat{\mathcal{H}}^{(\ee_x)}_{\text{surf}} (\kay_{\parallel}) = \begin{pmatrix}
		-\Delta_{\text{surf}}	&	-\Delta_1 (k_y - i k_z)	\\
		-\Delta_1 (k_y + i k_z)	&	\Delta_{\text{surf}}
	\end{pmatrix}	\\
	+ \Oh(\kay_{\parallel}^2).
\end{multline}
Up to $\Oh(\kay_{\parallel})$, the surface states resemble a two-component Dirac fermion moving in two spatial dimensions, with mass $\Delta_{\text{surf}}$ and ``speed of light" $\Delta_1$ (which has units of energy, since we work in unit where $a = 1$). As seen in Fig.~\ref{fig:zerofield}, the surface states at small momentum are well-approximated by the Dirac theory (red dashed curve), with $\Oh(\kay^2)$ Hamiltonian terms contributing only a small correction (black curve).

The leading-order massive-Dirac approximation of Eq.~\eqref{eq:surface_dirac} is particularly useful as a minimal model which can account for hinge modes. Consider a massive Dirac fermion with mass $\Delta_{\text{surf}}$. Jackiw and Rebbi \cite{JackiwRebbi} showed that gap-crossing chiral propagating bound states emerge along domain walls where the sign of $\Delta_{\text{surf}}$ changes. In the context of a 3D chiral hinge insulator, the hinges (where surfaces with $\en_1 \perp \en_2$ intersect) constitute such domain walls. However, we stress that the massive-Dirac picture should not be expected to accurately predict the dispersion of the surface states, because it does not fully account for the leading-order (in $\kay_{\parallel}$) contribution to their energy eigenvalues. The energy eigenvalues corresponding to Eq.~\eqref{eq:surface_dirac} are $E = \pm \sqrt{\Delta_{\text{surf}}^2 + \Delta_1^2 \kay_{\parallel}^2} \approx \pm \left( |\Delta_{\text{surf}}| + \Delta_1^2 \kay_{\parallel}^2 / |2 \Delta_{\text{surf}}| \right)$, where the approximation holds at small $\kay_{\parallel}$. Note that the surface state dispersion has no linear ($\Oh(\kay_{\parallel})$) term in the small-$\kay_{\parallel}$ limit, and so truncating the expansion of $\hat{\mathcal{H}}_{\text{surf}}$ at $\Oh(\kay_{\parallel})$ is not an order-consistent approximation. Especially in a magnetic field, Hamiltonian terms at $\Oh(\kay_{\parallel}^2)$ can become equally important to those at $\Oh(\kay_{\parallel})$. In particular, while their effect on the spectrum is quantitative (and small) in the absence of field, they lead to a particle-hole asymmetry in the case of non-zero field.  We address the $\Oh(\kay_{\parallel}^2)$ terms in section \ref{sec:k2_corrections}.

\subsection{Adding magnetic flux}	\label{subsec:dirac_landau}

\begin{figure}
	\centering
	\includegraphics[width=\linewidth]{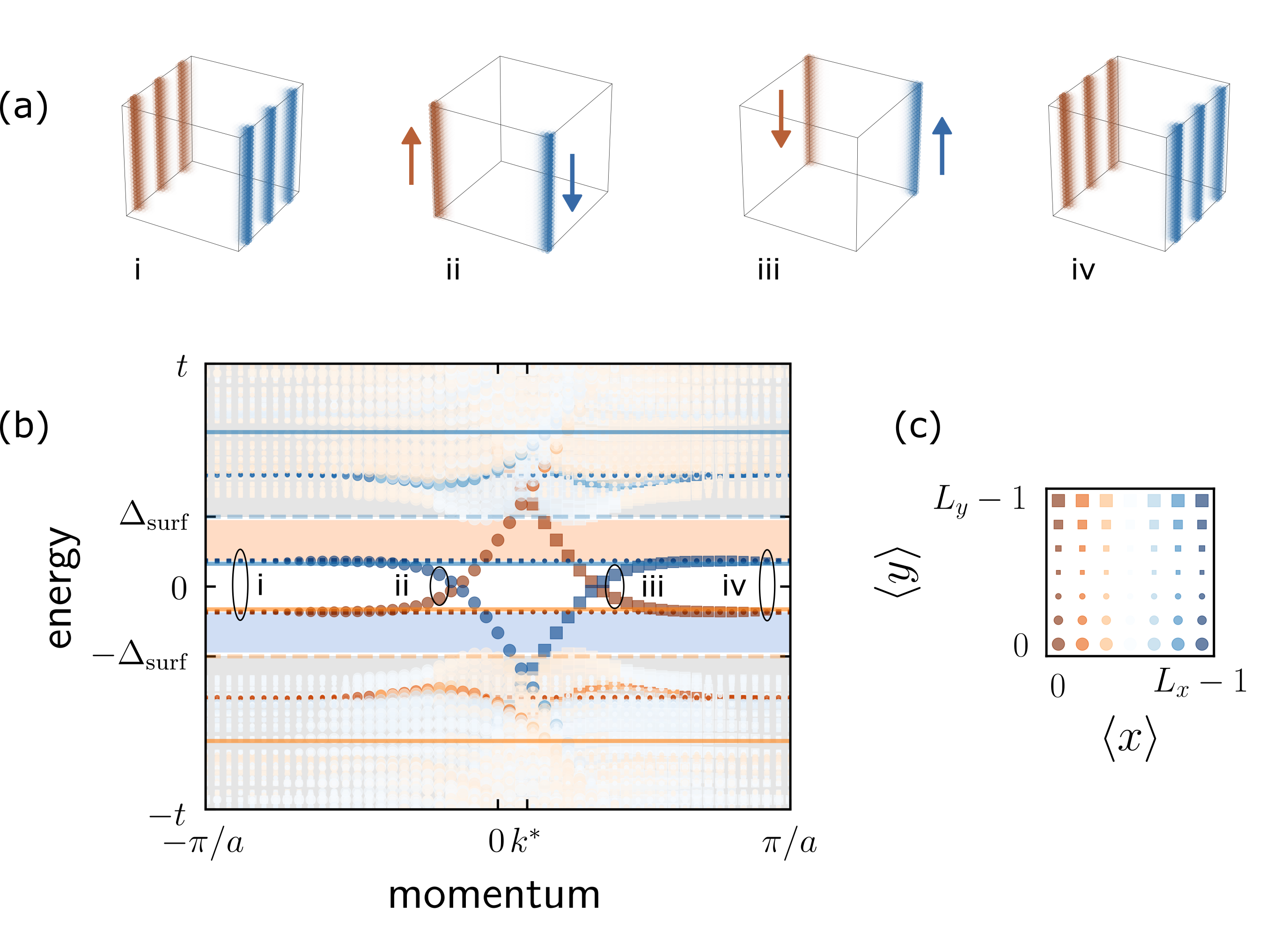}
	\caption{\label{fig:landau} Under magnetic field (with a flux per unit cell of $\Phi / \Phi_0 = 1/10$), the boundary of a 3D SOTI hosts Landau levels (i and iv) on the surfaces pierced by flux, 
		gapped dispersing states on the surfaces parallel to the flux, and chiral hinge modes (ii and iii) connecting the states on neighbouring surfaces.
		(a) Probability densities summed over each circled set of states in (b). (b) Energy eigenvalues of a sample with open boundary conditions in $x$ and $y$ ($L_x = L_y = 30$), and 
		with periodic boundary conditions in the $z$-direction (vertical in (a), out of the page in (c)).
		(c) Each energy eigenstate represented in (b) is labelled according to its expected position in the (cross-sectional) $xy$-plane.
		The orange (blue) dashed line indicates the lowest Landau level predicted to reside on 
		the left (right) surface of the sample according to the usual relativistic Landau theory. The orange (blue) solid lines indicate the predicted lowest and first 
		excited levels on the left (right) surface when $\Oh(\kay_{\parallel}^2)$ corrections are accounted for.}
\end{figure}

In units where $c = -\text{electron charge} = 1$, a magnetic gauge field $\Ay$ shifts the momentum operator $ \kay \rightarrow \kay - (q/c) \Ay = \kay + \Ay \equiv \boldsymbol{\pi}$, i.e.~$k_z \rightarrow k_z + B y$ in the Landau gauge. On surfaces $\en = \pm \ee_y$, which are parallel to the applied flux, the gauge field does not dramatically change the spectrum. Since the surface states are exponentially localized to the surfaces, $y$ is essentially constant for these states, and the gauge field simply shifts $k_z$ by the associated constant value of $A_z$. Further, the in-plane components of the gauge-invariant momentum $\boldsymbol{\pi}$ still commute with one another. On the other hand, on surfaces where $\en = \pm \ee_x$, the $y$- and $z$-components of $\boldsymbol{\pi}$ no longer commute. On the $\en = \ee_x$ surface, the Dirac Hamiltonian of Eq.~\eqref{eq:surface_dirac} becomes
\begin{equation}	\label{eq:dirac_landau_hamiltonian}
	\hat{\mathcal{H}}^{(\ee_x)}_{\text{surf}} = \begin{pmatrix}
		-\Delta_{\text{surf}}	&	\Delta_1 \sqrt{2 B} a	\\
		\Delta_1 \sqrt{2B} a^{\dagger}	&	\Delta_{\text{surf}}
	\end{pmatrix}
\end{equation}
with lowering operator $a \equiv i  \sqrt{\frac{B}{2}} \left(y + \frac{1}{B} (i k_y + k_z) \right)$.

An elementary calculation (e.g.~\cite{Levitan:SOTImagnetotransport}) gives the spectrum of relativistic Landau levels
\begin{subequations}		\label{eq:relativistic_landau}
	\begin{equation}	\label{eq:relativistic_landau_0}
		E_0 = \Delta_{\text{surf}}
	\end{equation}
	\begin{equation}	\label{eq:relativistic_landau_n}
		E_{n \ne 0} = \mathrm{sign}(n) \sqrt{\Delta_{\text{surf}}^2 + 2 B \Delta_1^2 |n|}.
	\end{equation}
\end{subequations}
States in the $n = 0$ level have the form
\begin{subequations}		\label{eq:relativistic_landau_states}
	\begin{equation}
		\ket{\psi_0} = \begin{pmatrix}
			0	\\
			\ket{0}
		\end{pmatrix},
	\end{equation}
	and states in the $n \ne 0$ levels are of the form
	\begin{equation}
		\ket{\psi_n} = \begin{pmatrix}
			u_n \ket{ |n| - 1}	\\
			v_n \ket{|n|}
		\end{pmatrix},
	\end{equation}
\end{subequations}
where $\lbrace \ket{n} \rbrace$ are the harmonic oscillator-like eigenfunctions of $a^{\dagger} a$.

The only spectral asymmetry in Eq.~\eqref{eq:relativistic_landau} is due to the lowest, $n = 0$ Landau level. Therefore, the Landau quantization on a single surface of a 3D SOTI is na\"{i}vely expected to display two qualitative features: (i) the $n=0$ Landau level exists away from zero energy, without a partner of opposite energy, and (ii) all other levels come in symmetric pairs around zero energy (the $\pm n^{th}$ levels). As shown in Fig.~\ref{fig:landau}, numerical diagonalization of the lattice model (Eq.~\eqref{eq:hamiltonian}) confirms the first of these qualitative predictions. Interestingly, the second prediction is violated --- on a given surface, the $n \ne 0$ levels are also asymmetric around zero energy \footnote{Indeed, there is no apparent $n = -1$ ($n = +1$) level on the blue $\en = \ee_x$ (orange $\en = -\ee_x$) surface in Fig.~\ref{fig:landau}(b).}. Quantitatively, the lowest level exists \textit{inside} the surface gap $\Delta_{\text{surf}}$ (dashed orange line in Fig.~\ref{fig:landau}(b)), as opposed to what would be expected from Eq.~\eqref{eq:relativistic_landau_0}.


As discussed in section \ref{sec:k2_corrections}, both of these deviations from Dirac behaviour --- the asymmetry of the $n \ne 0$ levels, and the $n = 0$ level falling inside the surface gap --- can be explained by including often-neglected Hamiltonian terms at $\Oh(\kay_{\parallel}^2)$. Including these terms yields significantly improved agreement with the numerical result for the lowest Landau level on each surface; the predicted lowest and first excited level on each surface are shown in solid lines in Fig.~\ref{fig:landau}(b), with the surface indicated by colour as throughout (see Fig.~\ref{fig:landau}(c)).
The energy of the first excited level deviates from its predicted value. This deviation is not surprising: the $|n| \ge 1$ levels are predicted at energy scales where the surface theory is no longer accurate even in zero field. Further, it is inevitable that higher-order-in-$\kay_{\parallel}$ terms (describing the short-distance physics) become increasingly important as the magnetic length $\sim 1/\sqrt{B}$ approaches the scale of the lattice spacing, since Hofstadter's butterfly differs drastically from a simple relativistic Landau fan. 

\subsection{Chiral hinge modes}	\label{subsec:hinge_theory}

\begin{figure}
	\centering
	\includegraphics[width=0.9\linewidth]{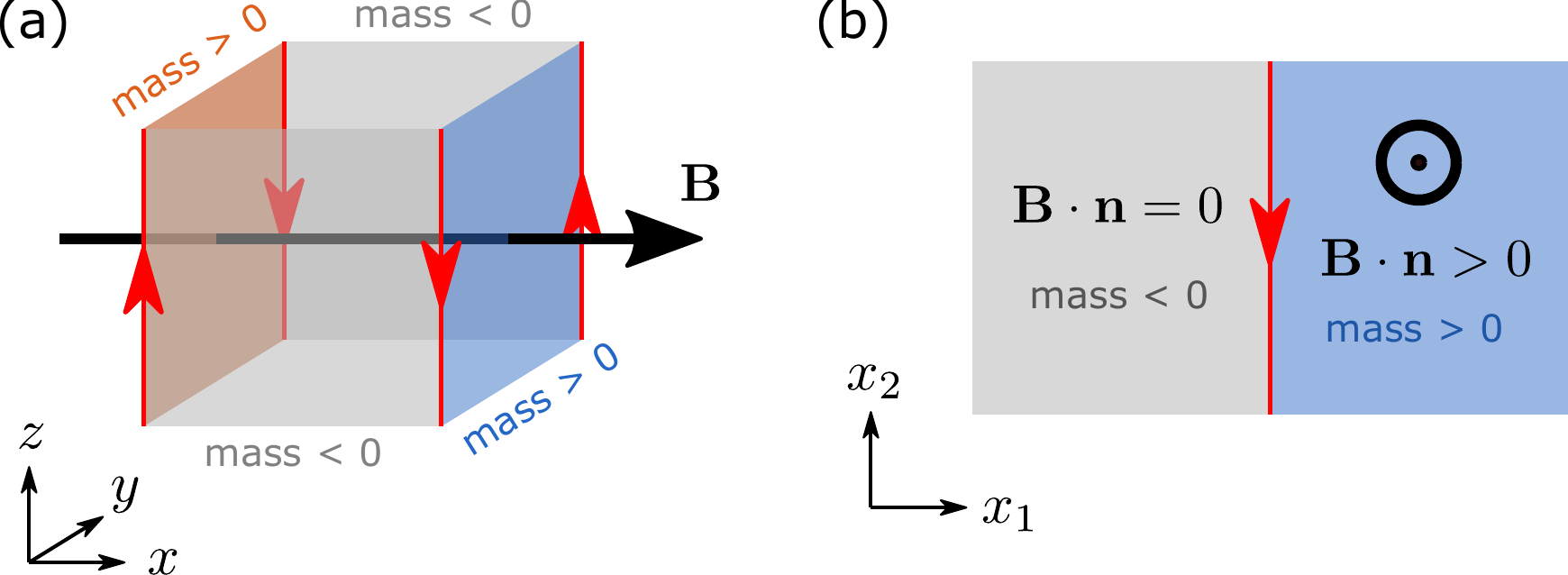}
	\caption{\label{fig:hinge_cartoon} (a) The chiral hinge modes of a 3D second-order topological insulator propagate along 1D (i.e.~codimension-2) submanifolds where the surface mass 
		gap changes sign. A magnetic field produces Landau levels on the surfaces pierced by flux (orange and blue). The in-gap hinge mode flows into the lowest Landau level as a
		function of momentum along the hinge (see Figs.~\ref{fig:landau}(b) and \ref{fig:hinge_spectrum}). 
		(b) A minimal 2D model containing a single hinge mode is obtained by ``unfolding" a hinge of the 3D model (see subsection \ref{subsec:hinge_theory}).}
\end{figure}

\begin{figure}
	\centering
	\includegraphics[width=0.8\linewidth]{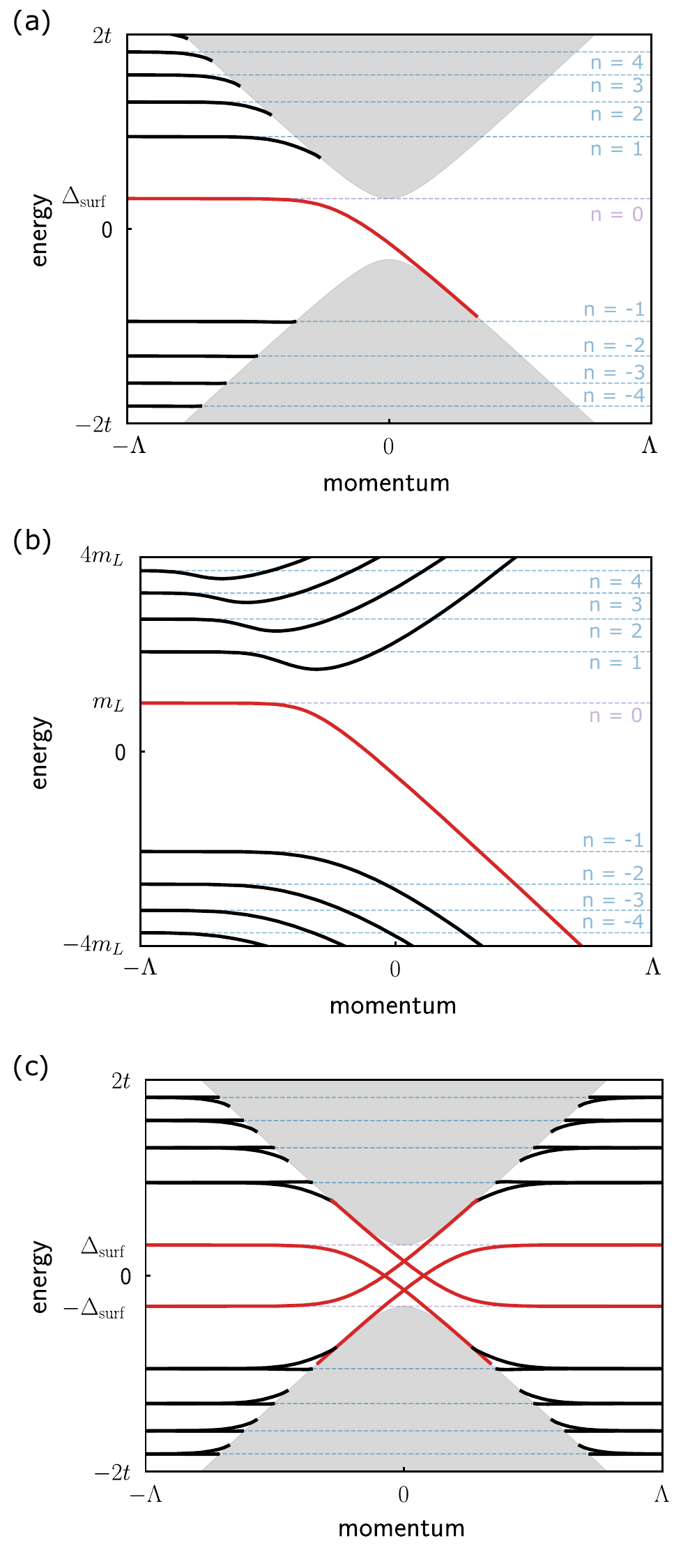}
	\caption{\label{fig:hinge_spectrum} The chiral Jackiw-Rebbi state which occurs at a domain wall in the sign of the Dirac mass survives when one half-plane is subject to a 
		magnetic field. This models the situation where a 3D second-order topological insulator (SOTI) is subject to a magnetic field $\Bee \parallel \ee_x$, piercing through two of the surfaces while parallel to the rest. We cut off the continuum theory at momentum $\Lambda = \pi / a$.
		As throughout, we use periodic boundary conditions along $z$ (the direction of the hinge).
		The solid lines in (a) and (b) are solutions to Eq.~\eqref{eq:hinge_matching}. Each of these states can be labeled by the 
		integer $n$ describing the relativistic (2D) surface Landau level to which they flow. The gray shaded regions indicate the Dirac cones on the zero field part of the plane.
		(a) corresponds to the $\Oh(\kay)$ boundary Dirac theory of the SOTI, with Dirac mass $\pm \Delta_{\text{surf}}$ changing only in sign across the domain wall. (b) shows the 
		result when, in addition, the mass on the Landau side is reduced to $m_{\text{Landau}} = \Delta_{\text{surf}} / 5$, bringing the first several levels inside the gap of the 
		Dirac side. (c) shows the analogous results to panel (a) for all four hinges of the SOTI.}
\end{figure}

In the absence of a magnetic field, the hinge modes of a second-order topological insulator can be understood as gapless modes pinned to the locations of sign changes in the mass gap of the insulator's surface states. The usual calculation in zero field involves matching decaying states on either side of the domain wall (i.e.~the hinge) \cite{JackiwRebbi}. In this subsection, we extend the usual calculation to include magnetic flux piercing through the surface on one side of the domain wall.

We consider the hinge between the $\en = - \ee_y$ and $\en = \ee_x$ surfaces, and take the hinge to reside at $x = y = 0$.
On the two surfaces in question ($\en \in \lbrace -\ee_y, \ee_x \rbrace$), the effective Hamiltonians are
\begin{subequations}
	\begin{equation}
		\hat{\mathcal{H}}_{\text{surf}}^{(-\ee_y)} \approx  - \Delta_1 \left( k_1 \hat{\rho}^{(-\ee_y)}_1 + k_2 \hat{\rho}^{(-\ee_y)}_2 \right)
			+ \Delta_{\text{surf}} \hat{\rho}^{(-\ee_y)}_3
	\end{equation}
	and
	\begin{equation}
		\hat{\mathcal{H}}_{\text{surf}}^{(\ee_x)} \approx  - \Delta_1 \left( k_1 \hat{\rho}^{(\ee_x)}_1 + k_2 \hat{\rho}^{(\ee_x)}_2 \right)
			- \Delta_{\text{surf}} \hat{\rho}^{(\ee_x)}_3.
	\end{equation}
\end{subequations}
$\hat{\rho}^{(\en)}_j$ acts like the $j^{th}$ Pauli matrix in the (two-band) space of the states on the $\en$ surface, with basis $\lbrace \ket{\psi^{(\en)}_{\en \cdot \boldsymbol{\sigma} = \pm1}} \rbrace$; note that different surfaces correspond to different two-dimensional subspaces of the bulk model's full four-band space. Here, $k_1 \equiv k_x$ ($k_y$) on the $\en = -\ee_y$ ($\en = \ee_x$) surface, while $k_2 \equiv k_z$ on both. 
The change in sign of the $\hat{\rho}^{(\en)}_3$ Dirac mass term (relative to the $\kay \cdot \hat{\boldsymbol{\rho}}^{(\en)}$ term) across the hinge gives rise to a chiral mode along the hinge.

In order to capture the essential physics of the sign change in the Dirac mass, we write down a minimal two-band model on a simplified geometry. We proceed by unfolding the hinge as shown in Fig.~\ref{fig:hinge_cartoon}(b), yielding a plane with coordinates $( x_1, x_2 )$. $x_1 < 0$ ($x_1 > 0$) corresponds to the bulk of the $\en = - \ee_y$ ($\en = \ee_x$) surface. The hinge lies along $x_1 = 0$. Before applying a magnetic field, the Hamiltonian is that of a Dirac fermion whose mass term changes sign at $x_1 = 0$, i.e.~$\mathcal{H} = - \left( \Delta_1 \kay \cdot \boldsymbol{\rho} + \text{sign}(x_1) \Delta_{\text{surf}} \rho_3 \right)$. Note that there is no longer a superscript $^{(\en)}$ on the Pauli matrices $\boldsymbol{\rho}$, which also no longer carry hats, since we have departed from the original four-band space of the three-dimensional bulk model.

In the bulk model, the magnetic gauge field $\Ay = (0, 0, By)$ is constant on the $\en = - \ee_y$ surface, which we can take to lie at $y = 0$ such that $\Ay$ vanishes there (for other choices of constant $y$ we could simply shift the gauge field). In the minimal hinge model, the gauge field translates to
\begin{equation}
	\Ay (x_1, x_2) = \begin{cases}
		(0, 0)			&	x_1 \le 0	\\
		(0, B x_1)		&	x_1 \ge 0
	\end{cases}
\end{equation}

The hinge and gauge field both break translation symmetry along $x_1$, but not along $x_2$. We therefore write the Hamiltonian in terms of position $x_1$ and momentum $k_2$. For $x_1 \le 0$,
\begin{subequations}	\label{eq:hinge_model}
	\begin{equation}
		\mathcal{H}_L = \begin{pmatrix}
			\Delta_{\text{surf}}			&	i \Delta_1	(\partial_1 + k_2)	\\
			-i \Delta_1 (- \partial_1 +  k_2)	&	-\Delta_{\text{surf}}
		\end{pmatrix},
	\end{equation}
	and for $x_1 \ge 0$,
	\begin{equation}
		\mathcal{H}_R = \begin{pmatrix}
			-\Delta_{\text{surf}}			&	\Delta_1 \sqrt{2B} a		\\
			\Delta_1 \sqrt{2B} a^{\dagger}	&	\Delta_{\text{surf}}
		\end{pmatrix},
	\end{equation}
\end{subequations}
where $a = i \sqrt{\frac{B}{2}} \left(x_1 + \frac{1}{B} (\partial_1 + k_2) \right)$.

On the left-hand side ($x_1 \le 0$), where there is no magnetic flux, we expect exponential decay away from the hinge, i.e.~$\ket{\psi_{\text{hinge}, L}} =  e^{\lambda x_1} \begin{pmatrix} \alpha_L	&	\beta_L \end{pmatrix}^T$ with $\lambda > 0$\footnote{Imaginary $\lambda$ would describe a propagating state. Also, $\lambda^2$ must be real to ensure real energy --- see Eq.~\eqref{eq:energy_lambda_relation}.}. The Schr\"{o}dinger equation corresponding to $\mathcal{H}_L$ then takes the form
\begin{equation}
	\begin{pmatrix}
		\Delta_{\text{surf}} - E	&	i \Delta_1 (k_2 + \lambda)	\\
		-i \Delta_1 (k_2 - \lambda)	&	- \Delta_{\text{surf}} - E
	\end{pmatrix}
	\begin{pmatrix}
		\alpha_L	\\
		\beta_L
	\end{pmatrix}
	= 0,
\end{equation}
implying
\begin{subequations}
	\begin{equation}	\label{eq:energy_lambda_relation}
	E^2 = \Delta_{\text{surf}}^2 + \Delta_1^2 (k_2^2 - \lambda^2)
	\end{equation}
	and
	\begin{equation}	\label{eq:dirac_matching}
		\frac{\alpha_L}{\beta_L} = \frac{i \Delta_1 (k_2 + \lambda)}{E - \Delta_{\text{surf}}}.
	\end{equation}
\end{subequations}
Eq.~\eqref{eq:dirac_matching} serves as a boundary condition for the other half-plane ($x_1 > 0$), to which we now turn our attention.

In the present context, following Ref.~\cite{Asaga:2020MagneticDirac}, it is convenient to write the lowering operator as $a = i (\partial_{\xi} + (\xi - \xi_0)/2)$, in terms of a rescaled coordiate $\xi = \sqrt{2B} x_1$ and $\xi_0 = - k_2 \sqrt{2/B}$. Squaring $\mathcal{H}_R$ gives
\begin{equation}
	\mathcal{H}_R^2
	= \begin{pmatrix}
		\Delta_{\text{surf}}^2 + 2 B \Delta_1^2 a a^{\dagger}	&	0	\\
		0	&	\Delta_{\text{surf}}^2 + 2 B \Delta_1^2 a^{\dagger} a
	\end{pmatrix}.
\end{equation}Acting with $\mathcal{H}_R^2$ on a state $\ket{\psi_{\text{hinge}, R}} = \begin{pmatrix} \alpha_R (\xi)	&	\beta_R (\xi) \end{pmatrix}^T$, the lower component satisfies
\begin{equation}	\label{eq:parabolic_cylinder_ode}
	\left[ \partial_{\xi}^2 + \frac{1}{2} + \nu - \frac{1}{4} (\xi - \xi_0)^2 \right] \beta (\xi) = 0,
\end{equation}
where $\nu = (E^2 - \Delta_{\text{surf}}^2)/(2B\Delta_1^2)$. Eq.~\eqref{eq:parabolic_cylinder_ode} is known as the parabolic cylinder ODE \cite{Asaga:2020MagneticDirac, Bateman:Transcendental}. It has two linearly-independent solutions; normalizability on the half-line $\xi > 0$ implies that the physical solution is the parabolic cylinder function $D_{\nu} (\xi - \xi_0)$ \footnote{If normalizability is instead demanded on the \textit{entire} real line (corresponding to an infinite, rather than semi-infinite plane), $\nu$ must take non-negative integer values $n$. In this case, the parabolic cylinder function $D_n$ coincides with the familiar harmonic oscillator wavefunction $\ket{n}$ up to a factor of $1/\sqrt{n!}$: $D_n (x) = 2^{-n/2} e^{-x^2/4} H_n (x/\sqrt{2})$, where $\lbrace H_n \rbrace$ are the Hermite polynomials.}. These functions satisfy the important relation $a D_{\nu} (\xi - \xi_0) = i \nu D_{\nu -1} (\xi - \xi_0)$. Turning attention back to $\mathcal{H}_R$ itself (instead of $\mathcal{H}_R^2$), it then follows that the state for $x_1 = \xi / \sqrt{2B} \ge 0$ is described by
\begin{equation}	\label{eq:landau_matching}
	\frac{\alpha_R (\xi)}{\beta_R (\xi)} = 
		\frac{\Delta_1 \sqrt{2B}}{E + \Delta_{\text{surf}}} \frac{(i \nu) D_{\nu - 1} (\xi - \xi_0)}{D_{\nu} (\xi - \xi_0)}.
\end{equation}
At the location of the hinge, the states $\ket{\psi_{\text{hinge}, L}}$ and $\ket{\psi_{\text{hinge}, R}}$ must match, so Eq.~\eqref{eq:dirac_matching} must agree with Eq.~\eqref{eq:landau_matching} evaluated at $\xi = 0$, i.e.
\begin{equation}	\label{eq:hinge_matching}
	\frac{\Delta_1 (k_2 + \lambda)}{E - \Delta_{\text{surf}}}
	= \frac{E - \Delta_{\text{surf}}}{\Delta_1 \sqrt{2B}} \frac{D_{\nu - 1} (k_2 \sqrt{2/B})}{D_{\nu} (k_2 \sqrt{2/B})}.
\end{equation} 
Together with Eq.~\eqref{eq:energy_lambda_relation}, which relates $E$ and $\lambda$, Eq.~\eqref{eq:hinge_matching} can be solved numerically to obtain the energy $E$ as a function of the momentum $k_2$ along the hinge. The resulting spectrum is shown in Fig.~\ref{fig:hinge_spectrum}(a). If the mass $\Delta_{\text{surf}}$ on the right (Landau) side of the problem is decreased to $m_{\text{Landau}} =  \Delta_{\text{surf}} / 5$, multiple Landau levels fit inside the Dirac gap, as shown in Fig.~\ref{fig:hinge_spectrum}(b). This spectrum resembles that described in Ref.~\cite{Asaga:2020MagneticDirac}, which considered the Dirac-Landau problem on the half-plane $x_1 \ge 0$ with an explicit boundary condition imposed along the edge at $x_1 = 0$. In our present context, the boundary condition is provided by matching with the gapped Dirac problem on the left half-plane. Notice that in both panels of Fig.~\ref{fig:hinge_spectrum}, at large negative momenta where the guiding centre falls deep in the right half plane ($\xi_0 \rightarrow + \infty$), the energies approach the familiar 2D bulk Landau levels.

\section{Boundary theory: beyond Dirac}	\label{sec:k2_corrections}
\subsection{Parabolic dispersion}	\label{subsec:zerofield_k2_corrections}
As previously described in subsection \ref{subsec:surface_dirac}, the 2D Dirac Hamiltonian $\sim - \Delta_1 \kay \cdot \boldsymbol{\rho} \pm \Delta_{\text{surf}} \rho_3$ has energy eigenvalues $E = \pm \sqrt{\Delta_{\text{surf}}^2 + \Delta_1^2 \kay^2}$, which have no linear-in-$\kay$ term when $\kay \rightarrow 0$. Therefore, truncating the Hamiltonian of Eq.~\eqref{eq:surface_dirac} at $\Oh(\kay_{\parallel})$ does not fully capture the leading-order dispersion; the gapped fermion on the surface of a chiral hinge insulator is not \textit{exactly} of Dirac character. 
Including terms at $\Oh (\kay_{\parallel}^2)$ in the surface Hamiltonian yields
\begin{widetext}
\begin{equation}	\label{eq:corrected_surface_hamiltonian}
	\hat{\mathcal{H}}^{(\ee_x)}_{\text{surf}} (\kay_{\parallel}) = \begin{pmatrix}
		-\Delta_{\text{surf}} - C (k_y^2 + \frac{1}{2} k_z^2)	&	-\Delta_1 (k_y - i k_z)	\\
		-\Delta_1 (k_y + i k_z)	&	\Delta_{\text{surf}} + C (k_y^2 + \frac{1}{2} k_z^2)
		\end{pmatrix}
	+ \Oh(\kay_{\parallel}^3)
\end{equation}
\end{widetext}
where $C = - \frac{t \Delta_2}{\sqrt{t^2 + \Delta_2^2}} = \Delta_{\text{surf}} t / \tilde{M}$. The terms $\propto C$ can be interpreted as an anisotropic $\kay_{\parallel}$-dependent contribution to the Dirac mass ($\rho_3$) term. The spectrum of this Hamiltonian is shown as the grey shaded region in Fig.~\ref{fig:zerofield}, bounded by the black curve corresponding to $k_y = 0$. Relative to the $\Oh (\kay)$ Dirac Hamiltonian, which would yield the red dashed curve in Fig.~\ref{fig:zerofield}, including the $\Oh (\kay^2)$ terms yields slightly improved agreement with our numerical lattice results.

\subsection{Landau levels: entering the butterfly's wing}	\label{subsec:landau_k2_corrections}
As in subsection \ref{subsec:dirac_landau}, we introduce a magnetic field via $\kay \rightarrow \kay + \Ay$ and work in terms of the relativistic Landau lowering operator $a \equiv i  \sqrt{\frac{B}{2}} \left(y + \frac{1}{B} (i k_y + k_z) \right)$. The Hamiltonian of Eq.~\eqref{eq:corrected_surface_hamiltonian} then becomes
\begin{widetext}
\begin{equation}	\label{eq:corrected_landau_hamiltonian}
	\hat{\mathcal{H}}^{(\ee_x)}_{\text{surf}} = \begin{pmatrix}
		-\Delta_{\text{surf}}		&	\Delta_1 \sqrt{\frac{4 \pi \Phi}{\Phi_0}} a	\\
		\Delta_1 \sqrt{\frac{4 \pi \Phi}{\Phi_0}} a^{\dagger}	&	\Delta_{\text{surf}}
	\end{pmatrix}
	- \frac{B C}{2} \left[ 3 ( a^{\dagger} a + 1/2) + \frac{1}{2} \left( a a + a^{\dagger} a^{\dagger} \right) \right] 
	\begin{pmatrix}
		1	&	0	\\
		0	&	-1
	\end{pmatrix}.
\end{equation}
\end{widetext}
The term $\propto a^{\dagger} a + 1/2$ does not spoil the general form of the Landau level (pseudospinor) solution given in Eq.~\eqref{eq:relativistic_landau_states}, $\ket{\psi_n} \propto \begin{pmatrix} u_n \ket{|n|-1} & v_n \ket{|n|} \end{pmatrix}^T$. On the other hand, the $a a + a^{\dagger} a^{\dagger}$ terms mix different functions $\ket{n}$. Their effect is therefore expected to be of higher order in perturbation theory, and we neglect these terms.
The resulting Landau level energies are
\begin{subequations}		\label{eq:corrected_landau_energies}
	\begin{equation}	\label{eq:corrected_lowest_level}
		E_0 = \Delta_{\text{surf}} + \frac{3 B C}{4}
	\end{equation}
	and
	\begin{multline}	\label{eq:corrected_higher_levels}
		E_{n \ne 0} = \frac{3 B C}{4} 	\\
		+ \text{sign}(n) \sqrt{2 B |n| \Delta_1^2 + (\Delta_{\text{surf}} + 3 B C |n| / 2)^2}.
	\end{multline}
\end{subequations}
Eq.~\ref{eq:corrected_landau_energies} explains several discrepancies between the familiar $\Oh(\kay)$ relativistic Landau picture and our numerical results for the 3D SOTI, shown in Fig.~\ref{fig:landau}(b) --- see Eq.~\ref{eq:relativistic_landau} and surrounding discussion. The $3 B C / 4$ term (arising from the ``zero point contribution" $1/2$ appearing in $a^{\dagger} a + 1/2 \sim (\kay + \Ay)^2$) is of particular importance. It accurately predicts the deviation of the lowest Landau level from $\Delta_{\text{surf}}$ (note that $C < 0 < B$ for the parameters we consider). As well, the same term in Eq.~\eqref{eq:corrected_higher_levels} predicts that on a given surface, the $n\neq 0$ levels (and not only the lowest level) are asymmetric around zero energy. This is in contrast to the relativistic Landau spectrum (Eq.~\eqref{eq:relativistic_landau}), where $E_n = - E_{-n}$ for $n \ge 1$. The $\Oh (\kay_{\parallel}^2)$ corrections yield an accurate prediction of the lowest Landau level away from the limit of vanishing flux, in the weak-but-finite flux regime of Fig.~\ref{fig:butterfly}. 

Note that the shift which leads to the asymmetry of the $n \ne 0$ levels vanishes when $B \rightarrow 0$. Even with the $\Oh(\kay_{\parallel}^2)$ terms, the zero-field Hamiltonian of Eq.~\eqref{eq:corrected_surface_hamiltonian} admits a particle-hole symmetry, which is broken by the magnetic field.

\subsection{$\Oh(\kay^2)$ terms in the BHZ model}

\begin{figure}[h!]
	\centering
	\includegraphics[width=0.9\linewidth]{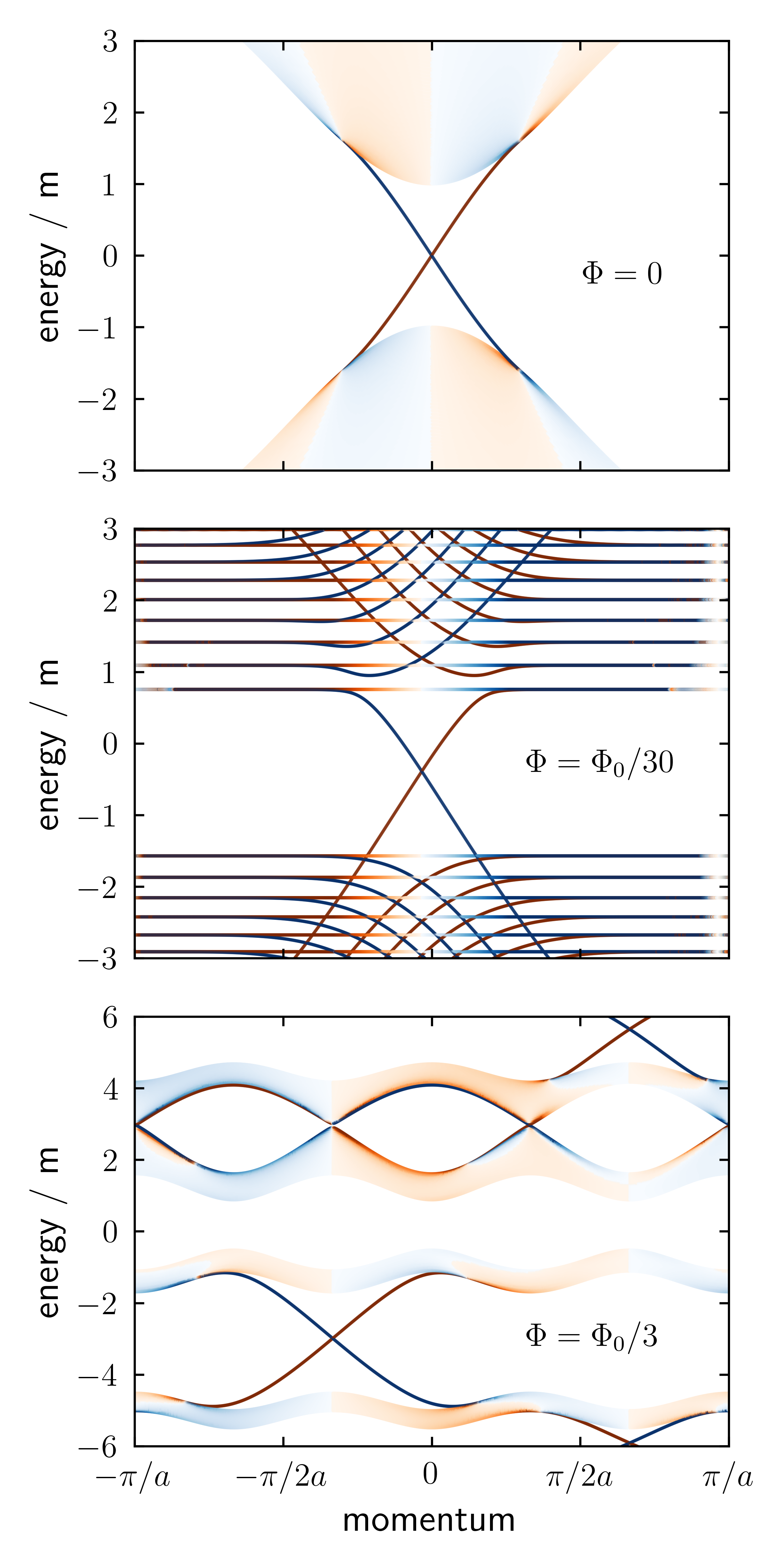}
	\caption{\label{fig:bhz_spectra} At low energies, in zero magnetic field (top panel), 
		each $\mathcal{T}$-broken (Chern) sector of the BHZ model for a 2D $\mathcal{T}$-symmetric topological insulator 
		resembles a massive Dirac fermion, and gives rise to chiral gap-crossing edge states (forming helical pairs when both sectors are considered). An applied perpendicular magnetic field gives rise to Landau levels when the 
		flux-per-plaquette is small (middle panel). As the applied field becomes large, the Landau levels give way to dispersing Chern bands 
		(bottom panel). Energies are coloured according to the average coordinate $\langle x \rangle$ of the associated states. 
		Dark orange (blue) indicates an edge state localized near the left-hand (right-hand) side of the sample. Lighter colours indicate 2D bulk states.}
\end{figure}

In the previous subsection, we described discrepancies between the surface states of a second-order topological insulator and the continuum massive Dirac model. Naturally, due to band curvature, \textit{any} lattice realization of a gapped Dirac fermion will deviate from the ideal continuum case, especially away from the Dirac point. As a concrete example, we consider the 2D BHZ model \cite{BernevigHughesZhang}, describing a 2D $\mathcal{T}$-symmetric topological insulator. The BHZ model has two $\mathcal{T}$-broken sectors (of opposite Chern number) which map onto one another under $\mathcal{T}$. In momentum space, one $\mathcal{T}$-broken sector is described by
\begin{subequations}		\label{eq:bhz}
	\begin{equation}	
		\mathcal{H} (\kay) = \epsilon (\kay) \sigma_0 + \boldsymbol{d} (\kay) \cdot \boldsymbol{\sigma}
	\end{equation}
	where
	\begin{equation}
		d_1 + i d_2 = \alpha (\sin k_x + i \sin k_y),
	\end{equation}
	\begin{equation}
		d_3 = m + 2 \beta (\cos k_x + \cos k_y - 2),
	\end{equation}
\end{subequations}
and we take $\epsilon (\kay) = 0$ for simplicity. 
When $d_3 = 0 \; \forall \; k$, the Hamiltonian of Eq.~\eqref{eq:bhz} has Dirac nodes at all time-reversal invariant momentum (TRIM) points; $d_3(k)$ serves 
as a mass term which opens a gap at these points. In order to make connection with the SOTI surface described earlier in this work, we will choose parameters such that the 
gap at the $\Gamma$ point is much smaller than that at the other TRIM points. Therefore, at low energy, we have a single massive Dirac point.

As in the case of the 3D SOTI, a perpendicular magnetic field $\Bee = B \ee_3$ can be included via Peierls substitution in real space: $c_{\ar + \ee_j}^{\dagger} c_{\ar} \rightarrow e^{i (q / \hbar) \int_{\ar}^{\ar + \ee_j} \mathrm{d} \boldsymbol{\ell} \cdot \Ay} c_{\ar + \ee_j}^{\dagger} c_{\ar}$, where $\nabla \times \Ay = \Bee$. We choose the Landau gauge $\Ay = (0, B x_1)$, which preserves translation invariance along $x_2$. The results of numerical diagonalization on a periodic strip of finite width ($L_x = 90$) are shown in Fig.~\ref{fig:bhz_spectra}. The upper panel corresponds to zero magnetic field, and shows the expected gapped dispersion resembling a massive Dirac cone, with edge states crossing the gap. The middle panel corresponds to a flux-per-plaquette of $\Phi / \Phi_0 = 1 / 30$. In this regime of moderate magnetic field, the spectrum resembles that of Fig.~\ref{fig:hinge_spectrum}, or that discussed in Ref.~\cite{Asaga:2020MagneticDirac}, consisting of Landau levels and edge states. The lower panel corresponds to $\Phi / \Phi_0 = 1/3$. Such a strong magnetic flux puts the model deep into the Hofstadter regime, where Landau levels give way to dispersing Chern bands. In all three panels, the model parameters are $\alpha / m = 2$ and $\beta / m = 1.2$.

\begin{figure}[t]
	\centering
	\includegraphics[width=0.9\linewidth]{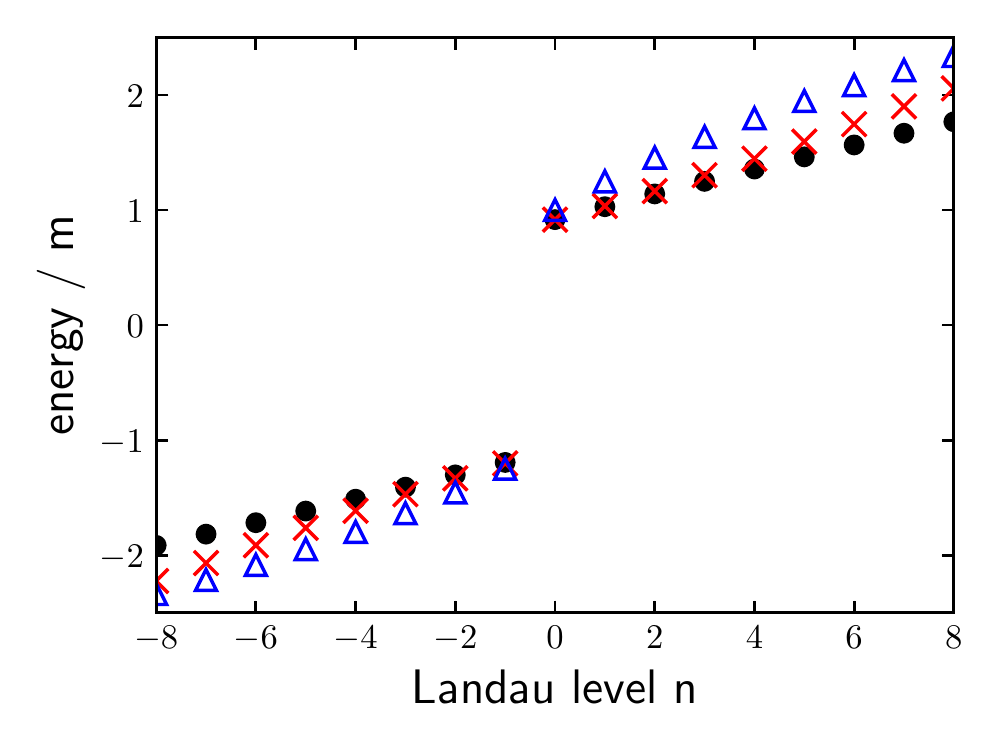}
	\caption{\label{fig:bhz_landau} When a magnetic field (with $\Phi / \Phi_0 = 1/90$) is applied to one $\mathcal{T}$-broken sector of the BHZ model, Landau levels emerge (black dots). These levels deviate from the na\"{i}ve expectation of relativistic Landau quantization (blue triangles). The lowest levels near zero energy are accurately predicted by the corrected continuum theory at $\Oh(\kay^2)$ (red crosses).}
\end{figure}

In the continuum limit ($\kay \rightarrow 0$), an analytic treatment of the BHZ model in a perpendicular magnetic field can be performed in direct analogy to subsection \ref{subsec:landau_k2_corrections}. One can expand $\boldsymbol{d}$ to $\Oh(\kay^2)$ and then make the minimal substitution $\kay \rightarrow \kay + \Ay$. Note that in the case of the BHZ model, the dispersion is isotropic at small $\kay$, and $aa + a^{\dagger} a^{\dagger}$ terms do not arise. Fig.~\ref{fig:bhz_landau} compares the resulting Landau level energies (red crosses) against the results of numerical diagonalization of the lattice BHZ model, with magnetic field included via Peierls substitution (black dots). The numerical data correspond to model parameters as in Fig.~\ref{fig:bhz_spectra}, with periodic boundary conditions in both directions, and with magnetic flux-per-plaquette $\Phi / \Phi_0 = 1/90$. In this weak-field regime, the $\Oh (\kay^2)$ theory accurately predicts the first few positive- and negative-energy Landau levels closest to zero energy.

\section{Conclusion}
Using a 2D surface theory constructed from a 3D lattice model, we have described the chiral metallic 1D hinge states and gapped 2D surface states of a 3D second-order topological insulator (SOTI) subject to an applied magnetic gauge field. By generalizing the Jackiw-Rebbi domain wall to include magnetic flux through one half-plane, we have modelled the situation where the magnetic field pierces through two surfaces of the 3D SOTI and is perpendicular to the other two. The resulting domain wall theory accurately captures several qualitative features of the numerical 3D lattice model, including spectral flow between the lowest Landau level on one surface and the massive Dirac cone on its neighbour, mediated by a chiral 1D mode. Our approach could be straightforwardly extended to allow for flux through both half-planes, modelling the situation where the magnetic field is not parallel to any surface.

As the field is increased, the lattice Landau level structure on surfaces pierced by flux deviates from the familiar result expected from a massive Dirac fermion. We have shown that for a range of magnetic fields, the discrepancy is the result of often-neglected Hamiltonian terms in the surface theory at $\Oh(\kay^2)$. Most importantly, these terms lead to an overall shift of the Landau level spectrum which is linear in the field $B$, breaking the symmetry of the excited Landau levels ($|n| \ge 1$) around zero energy. The importance of the $\Oh(\kay^2)$ terms is not specific to the 3D SOTI model we considered: because the dispersion of a massive Dirac fermion is quadratic at small momentum, such terms are generically expected to become relevant in \textit{any} lattice model containing a 2D massive Dirac fermion. Away from the Hofstadter regime, the Landau level shift is therefore expected. As an example, we showed how the $\Oh (\kay^2)$ continuum theory accurately predicts the Landau level structure of a 2D BHZ model pierced by magnetic field.

\acknowledgments{
The authors thank E.~Altman, A.~MacDonald, R.~Queiroz, and A.~Stern for useful and enlightening discussions. The work in this manuscript has been supported by the Natural Sciences and Engineering Research Council of Canada (NSERC) (TPB) and by the Fonds de recherche du Québec – Nature et technologies (FRQNT) (BAL and TPB).
}

\bibliography{hoti}

\end{document}